\documentclass[twocolumn,showpacs]{revtex4}

\usepackage{graphicx}
\usepackage{dcolumn}
\usepackage{bm}

\input epsf

\begin{document}

\title{\Large Schwarzschild Solution on the Brane}

\author{\bf~Subenoy Chakraborty\footnote{schakraborty@math.jdvu.ac.in}
and~Tanwi~Bandyopadhyay}

\affiliation{Department of Mathematics, Jadavpur University,
Kolkata-32, India.}

\date{\today}

\begin{abstract}
In this communication we have shown that Schwarzschild solution is
possible in brane world for some specific choices of brane matter
and the non local effects from the bulk. A conformally flat bulk
space time with fine-tuned vacuum energy (brane tension) shows that,
Schwarzschild solution may also be the vacuum solution for brane
world scenario.
\end{abstract}


\maketitle

In brane world scenario, our universe is a 3-brane [1] embedded in a
five dimensional bulk. In fact, the 3-brane is a domain wall
separating two semi-infinite anti de-Sitter regions. The standard
model fields are confined to the brane while gravity propagates in
the surrounding bulk. The field equations on the brane are derived
[2,~3] from the Einstein equations on the bulk, using the
Gauss-Codazzi equations and the matching conditions with $Z_{2}$
symmetry. The effective Einstein equations on the brane has the form

\begin{equation}
G_{\mu\nu}=-\Lambda~g_{\mu\nu}+\kappa^{2}T_{\mu\nu}+\frac{6\kappa^{2}}
{\lambda}~\Pi_{\mu\nu}-E_{\mu\nu}
\end{equation}

with $\Lambda=\frac{1}{2}(\Lambda_{5}+\kappa^{2}\lambda)$.\\

where $\Lambda$ and $\Lambda_{5}$ are the cosmological constants on
the brane and in the bulk respectively and $T_{\mu\nu}$ is the
energy momentum tensor for the matter in the bulk. Compared to
Einstein's general relativity, there are two additional terms in the
energy momentum tensor due to embedding of the brane into the bulk.
The term $\Pi_{\mu\nu}$ is quadratic in $T_{\mu\nu}$, arising from
the extrinsic curvature terms in the projected Einstein tensor and
is given by

\begin{equation}
4\Pi_{\mu\nu}=\frac{1}{3}TT_{\mu\nu}-T_{\mu\alpha}T_{\nu}^{\alpha}
+\frac{1}{2}g_{\mu\nu}\left(T_{\alpha\beta}T^{\alpha\beta}-\frac{1}{3}T^{2}\right)
\end{equation}

with $T=T_{\alpha}^{\alpha}$. This is known as local bulk effects.\\

The second correction term $E_{\mu\nu}$ stands for non local bulk
effects and is the projection of the five dimensional Weyl tensor
$~^{(5)}C_{ABCD}$ onto the brane, viz.

\begin{equation}
E_{\mu\nu}=~^{(5)}C_{ABCD}~\delta_{\mu}^{A}\delta_{\nu}^{C}n^{B}n^{D}
\end{equation}

where $n^{A}$ is the unit normal to the brane and $E_{\mu}^{\mu}=0$
(traceless).\\

When matter on the brane is absent (i.e, $T_{\mu\nu}\equiv0$) and
the brane tension is fine tuned so that, $\Lambda=0$, then the above
modified Einstein equation (1) simplifies to

\begin{equation}
G_{\mu\nu}=-E_{\mu\nu}
\end{equation}

This is termed as vacuum Einstein equations on the brane.
Subsequently, Dadhich et al.[4] has obtained Reissner-N\"{o}rdstrom
black hole solution but without electric charge being present. It is
interpreted as tidal charge, arising from the projection onto the
brane of free gravitational field effects in the bulk. Vollick [5]
has obtained some solutions solving $~^{(4)}R=0$. He speculated that
the Weyl term could be responsible for the observed dark matter in
the universe. Then Bronnikov and Kim [6] have presented static,
spherically symmetric Lorentzian wormhole solutions by solving
$~^{(4)}R=0$.\\

In the present work, we consider a static spherically symmetric
brane model with isotropic fluid confined in the brane. So, in an
orthonormal reference frame, the form of the energy momentum tensor
is

\begin{equation}
T_{\mu\nu}=diag(\rho,p,p,p)
\end{equation}

Further, due to static spherically symmetric nature of the problem,
the projected Weyl tensor has the form [7]

\begin{equation}
E_{\mu\nu}=diag~[\epsilon(r),\sigma_{r}(r),\sigma_{t}(r),\sigma_{t}(r)]
\end{equation}

Then the modified Einstein equations on the brane are

\begin{equation}
\left.
\begin{array}{ll}
G_{tt}=8\pi~G\rho^{eff}\\\\
G_{rr}=8\pi~Gp_{r}^{eff}\\\\
G_{\theta\theta}=8\pi~Gp_{t}^{eff}\\\\
\end{array}
\right\}
\end{equation}

\newpage
where

\begin{equation}
\left.
\begin{array}{ll}
\rho^{eff}=\rho\left(1+\frac{\rho}{2\lambda}\right)-\frac{\epsilon}{8\pi}\\\\
p_{r}^{eff}=p\left(1+\frac{\rho}{\lambda}\right)+\frac{\rho^{2}}{2\lambda}
-\frac{\sigma_{r}}{8\pi}\\\\
p_{t}^{eff}=p\left(1+\frac{\rho}{\lambda}\right)+\frac{\rho^{2}}{2\lambda}
-\frac{\sigma_{t}}{8\pi}\\\\
\end{array}
\right\}
\end{equation}

Now if the static spherically symmetric space time in the brane is
chosen to be Schwarzschild, then $G_{tt}=0=G_{rr}=G_{\theta\theta}$
and consequently we have

\begin{equation}
\left.
\begin{array}{ll}
\rho\left(1+\frac{\rho}{2\lambda}\right)-\frac{\epsilon}{8\pi}=0\\\\
p\left(1+\frac{\rho}{\lambda}\right)+\frac{\rho^{2}}{2\lambda}
-\frac{\sigma_{r}}{8\pi}=0\\\\
p\left(1+\frac{\rho}{\lambda}\right)+\frac{\rho^{2}}{2\lambda}
-\frac{\sigma_{t}}{8\pi}=0\\\\
\end{array}
\right\}
\end{equation}

with

\begin{equation}
-\epsilon+\sigma_{r}+2\sigma_{t}=0
\end{equation}

Thus we have four algebraic equations containing five unknowns viz.
$\rho$, $p$, $\epsilon$, $\sigma_{r}$ and $\sigma_{t}$. So, we have
the freedom to choose one more relation among these unknowns.
However, from the second and third equations of (9), one gets
$\sigma_{r}=\sigma_{t}(=\sigma)$ and hence essentially we have the
following two equations containing three unknowns

\begin{equation}
\left.
\begin{array}{ll}
\rho\left(1+\frac{\rho}{2\lambda}\right)-\frac{3\sigma}{8\pi}=0\\\\
\text{and}~~~~p\left(1+\frac{\rho}{\lambda}\right)+\frac{\rho^{2}}{2\lambda}
-\frac{\sigma}{8\pi}=0\\\\
\end{array}
\right\}
\end{equation}

We shall now present solutions for the following cases:\\

\textbf{\underline{Case-I:}}~~~~~~~~~~~~~\underline{$\sigma=0$}\\

Then we have two possibilities\\

~~~~~~~$\rho=0,~~p=0$~~~ or ~~~$\rho=-2\lambda,~~p=2\lambda$.\\

Thus Schwarzschild solution is possible for either vacuum brane or
brane with strange matter distribution ($\rho<0$ and $p>0$) embedded
in a conformally flat bulk.\\

\textbf{\underline{Case-II:}}~~~~~~~~~~~~~\underline{$p=0$ (\bf{Dust})}\\

Then we have ~~~~~$\rho=\lambda$ ~~~and ~~~$\sigma=4\pi\lambda$\\.

So it is possible to have Schwarzschild solution for dust brane with
non local correction term behaving as perfect fluid with radiation
equation of state.\\

\textbf{\underline{Case-III:}}~~~~~~~~~~~~~\underline{$p=\omega\rho$
(\bf{Perfect fluid})}\\

The expressions for $\rho$ and $\sigma$ are given by\\

~~~~~~~~~~~~~~~~$\rho=\lambda\left(\frac{1-3\omega}{1+3\omega}\right)$\\

and~~~~~~~~~~~$\sigma=4\pi\lambda\frac{(1-3\omega)(1+\omega)}{(1+3\omega)^{2}}$\\

with ~~~~~~$-1/3<\omega<1/3$.\\

So, a realistic perfect fluid with linear equation of state has
Schwarzschild solution provided that the non local bulk effect
also behaves as a perfect fluid. Therefore by fixing the geometry
of the brane world, it is possible to determine various
possibilities for the matter to have prescribed geometric
structure. Usually in the literature, matter is prescribed and we
have to determine the geometry by solving second order non linear
differential equations. On the other hand, in the present work,
we have fixed the geometry and matter is determined by solving a
set of algebraic equations. So we may conclude that, our method
is more suitable than the existing standard approach. Lastly, we
conclude that Birkhoff's theorem for unique spherically symmetric
solution is not possible in brane scenario.\\

{\bf Acknowledgement}:\\

The work has been done during a visit to IUCAA under the
associateship programme. The authors gratefully acknowledge the warm
hospitality and facilities of work at IUCAA. Also T.B is thankful to
CSIR, Govt. of India, for awarding Junior Research Fellowship.\\

\end{document}